\documentclass{article}

\usepackage{PRIMEarxiv}

\usepackage[utf8]{inputenc} 
\usepackage[T1]{fontenc}    
\usepackage{hyperref}       
\usepackage{url}            
\usepackage{booktabs}       
\usepackage{amsfonts}       
\usepackage{nicefrac}       
\usepackage{microtype}      
\usepackage{lipsum}
\usepackage{fancyhdr}       
\usepackage{graphicx}       
\graphicspath{{media/}}     
\usepackage{amsmath}
\usepackage{tabularray}
\usepackage{adjustbox}
\usepackage{pgffor}  
\usepackage{tabularx} 
\usepackage{threeparttable}
\usepackage{booktabs,caption}
\usepackage{algpseudocode}
\usepackage{algorithm}
\title{Exploring Public Opinion on Responsible AI Through The Lens of Cultural Consensus Theory}

\author{Necdet Gürkan \\
 The School of Business \\
 Stevens Institute of Technology \\
 {\underline{ ngurkan@stevens.edu}} \\ \\ \And
 Jordan W. Suchow \\
 The School of Business \\
 Stevens Institute of Technology \\
 {\underline{ jws@stevens.edu} } \\
 }

\date{}

\begin{document}

\maketitle

\begin{abstract}

As the societal implications of Artificial Intelligence (AI) continue to grow, the pursuit of responsible AI necessitates public engagement in its development and governance processes. This involvement is crucial for capturing diverse perspectives and promoting equitable practices and outcomes. We applied Cultural Consensus Theory (CCT) to a nationally representative survey dataset on various aspects of AI to discern beliefs and attitudes about responsible AI in the United States. Our results offer valuable insights by identifying shared and contrasting views on responsible AI. Furthermore, these findings serve as critical reference points for developers and policymakers, enabling them to more effectively consider individual variances and group-level cultural perspectives when making significant decisions and addressing the public's concerns.

\end{abstract}

\subsubsection*{Keywords:}

artificial intelligence, public perception, responsible AI, cultural consensus theory, Bayesian modeling

\section{Introduction}
Artificial Intelligence (AI) is a rapidly advancing field of science and technology that involves the use of algorithms and computer systems to mimic intelligent behavior \cite{russell2009artificial}. Advances in AI have impacted nearly all aspects of society, including the labor market, transportation, healthcare, education, virtual environments, and national security \cite{NationalArtificialIntelligence}. 
Applications of AI in these various domains are continually being developed and actively incorporated into daily life, potentially effecting transformative changes in our society. While technology developers and policymakers have begun to more frequently discuss the societal implications and responsibilities of AI, civil society groups, particularly underrepresented ones, argue that the public should play a role in shaping AI technology developments \cite{west2019discriminating}. Consequently, the concept of responsible AI has emerged as a pivotal principle guiding the governance of AI systems \cite{jobin2019global}.

As AI technology continues to advance and spark discussions, it is crucial to recognize that AI systems are imperfect. They can fail to meet their intended goals for various reasons, such as biased training data or the unintended and negative consequences of their recommendations and decisions \cite{crawford2016there, mittelstadt2016ethics, gurkan2021evaluation}. Furthermore, the design and implementation of these systems can often reflect and perpetuate societal biases, underlining the importance of continuous scrutiny and ethical considerations in AI development and deployment. Given that AI's influence has been noted in justice, privacy, stock and commodity trading, labor markets, and even democratic election outcomes \cite{bankins2022ai, calo2017artificial, samtani2023deep}, the White House Science and Technology Policy (2020) started a project called American AI Initiative to stress the importance of the government investing money in AI technologies and earning people's trust in the government's use of AI. Meanwhile, members of the U.S. Congress proposed the Algorithmic Accountability Act (2022), a law mandating that companies  examine their automated decision-making systems for potential risks, bias, unfairness, and privacy threats, thus promoting responsible AI use. 

Private corporations have proposed the idea of ``extended corporate citizenship'' \cite{matten2005corporate}, wherein ethical businesses work collaboratively with local actors, governments, and civil society to encourage responsible innovation processes \cite{adams2006innovation, stafford2017integration}. This body of work, implicitly and explicitly, often leans on the principles of ``deep democracy'' \cite{michelman2018can}, underscoring the value of open engagement processes based on principled communication among a diverse set of stakeholders. This implies that innovators, by actively engaging in public debates and discourse, can play a significant role in promoting responsible innovation.

As both government and private corporation actors start taking steps to develop more responsible AI, it is becoming increasingly important for developers, policymakers, and risk communicators to gain an empirical understanding of the diverse attitudes that various segments of the public have towards technology. Why does this matter? The primary reason is that when the public participates in decision-making about science and technology, it can lead to enhanced accountability, better decisions, and improved trust in those making policy decisions \cite{holdren2011principles}. However, to inspire a wide range of people to participate in these efforts, scientific communication must address their specific concerns \cite{scheufele2021we}. Therefore, public involvement is crucial for guiding the direction of AI research and usage \cite{buhmann2021towards}. 

The process of more responsible AI begins with gaining a deep, factual understanding of the varied perspectives of the public, as perceptions of responsible AI can greatly vary depending on individual beliefs, personal experiences, subject familiarity, and cultural constructs \cite{jakesch2022different}. These elements potentially shape the shared knowledge or beliefs within the population, thereby revealing the collective thinking or beliefs of a specific group. Yet, while researchers have noted these individual and group-level differences in perceptions of responsible AI, there remains a shortfall in studies that probe into which specific subjects contribute to these variances. Such research could offer invaluable insights for developers and policymakers, enabling them to address divergent opinions and bolster consensus beliefs. By dealing with the most contentious subjects, we could thereby cultivate a more comprehensive understanding of prevailing perceptions within a particular culture or community about responsible AI, ultimately enhancing the perception of emerging technologies as responsible and promoting their broader adoption within society.

Cultural Consensus Theory (CCT) is a statistical framework that can be used to infer the cultural beliefs influencing social practices and the degree to which individuals know or express those beliefs \cite{romney1986culture}. These models provide an opportunity to study individual differences in whether a group member conforms to the consensus in a community, and they allow people to differ in both their level of cultural knowledge and response biases. Researchers have applied the CCT framework to find practical and concise definitions of beliefs that are accepted by a group sharing common knowledge. These models have been widely used to study, healthy food beliefs \cite{gurkan2023food}, cognitive evaluation \cite{heshmati2019does}, and online communities \cite{gurkan2022learning}.

CCT offers a useful tool to identify consensus beliefs, variations in beliefs between and within groups, and how understanding such beliefs can benefit policymakers and developers. In this study, we applied the CCT to understand the diverse beliefs about responsible AI within society, using a nationally representative survey dataset. However, CCT has limitations. It assumes a finite-dimensional representation that characterizes data, which can be ineffective in a sparse data regime. To address this, we introduce the infinite Cultural Consensus Theory (iCCT), a Bayesian non-parametric model that enhances CCT by drawing cultures from a Dirichlet Process using stick-breaking construction. By applying iCCT to the public's perception of responsible AI, we aim to answer the following questions:

\begin{enumerate}
\item Is there a consensus among the U.S. population about responsible AI?
\item If not, which aspects of responsible AI exhibit the least consensus between and within groups?
\item How can a consensus model assist developers and policymakers in the context of responsible AI?
\end{enumerate}

\section{Related Work}

The concept of responsible AI has emerged as a guiding principle for the governance of AI systems \cite{jobin2019global}. This framework is intended to ensure that AI technologies are harnessed, deployed, evaluated, and monitored in a manner that creates new opportunities for improved service provision, while simultaneously adhering to ethical standards \cite{wang2020toward}. Central to this approach is the design and implementation of AI solutions that are not only ethical and transparent, but also accountable - aspects crucial in maintaining trust among individuals and minimizing invasions of privacy \cite{samtani2023deep}. By placing humans, particularly end-users, at its core, responsible AI seeks to meet stakeholder expectations, as well as comply with relevant regulations and laws.

Most research on the responsibility gap has been normative, prescribing ethical principles and proposing solutions. However, there is a growing need for practical and proactive guidelines; as Mittelstadt puts it, “principles alone cannot guarantee ethical AI” \cite{mittelstadt2019principles}. Some even argue that normative approaches can be inappropriate as they may hinder AI’s adoption in the long run \cite{bonnefon2020moral}. In contrast, relatively little attention has been paid to understanding the public’s views on this issue, despite them likely being the most affected stakeholders when AI systems are deployed \cite{rahwan2018society}.

Private companies, public sector organizations, and academic groups have published ethics
guidelines with values they consider important for responsible AI \cite{jobin2019global}. While these values preferences heavily depend on people's background, experiences, and beliefs, AI technologies are often developed by relatively homogeneous and demographically skewed subsets of the population \cite{mikalef2022thinking}. Given the lack of reliable data on
other groups’ priorities for responsible AI, practitioners may unknowingly encode their own biases and assumptions into their concept and operationalization of responsible AI \cite{raji2020closing}. \cite{jakesch2022different} surveyed the US population, crowdworkers, and AI practitioners and demonstrated that AI practitioners’ value priorities differ from those of the general public. 

Engaging the public effectively with complex technologies necessitates a nuanced understanding of how diverse audiences comprehend and communicate about disruptive technologies that carry significant social implications. Researchers have been shown that segmentation analysis can be an effective tool for understanding and addressing the needs of diverse communities, and it could add nuance to our understanding of public support for AI in the United States \cite{bernstein2018beyond}. A 2018 survey showed varied attitudes toward AI development: 41\% of U.S. adults showed some level of support, 28\% were neutral, and 22\% opposed it \cite{zhang2019artificial}. Most respondents, however, expressed the need for careful management of AI and robots, particularly concerning potential violations of civil liberties \cite{zhang2019artificial}. However, dichotomizing public responses to new technologies as simply being 'for' or 'against' is, at best, overly simplistic and, at worst, potentially misleading.

Adopting simplistic or one-size-fits-all messaging approaches has been shown to be less effective than communication strategies tailored to address audiences' specific concerns (Tallon, 2000). Bao et al. (2022) proposed using Latent Class Analysis (LCA) to segment the U.S. population based on their perceptions of the possible risks and benefits of AI. They found that U.S. adults do not form a monolith in terms of their attitudes towards AI, and cannot be dichotomously segmented as perceiving AI as either 'risky' or 'beneficial.' While LCA analysis effectively uncovers diverse attitudes and opinions regarding AI, it overlooks the concept that knowledge and beliefs about AI are socially distributed and reflect the cultural norms and values of a society. Furthermore, LCA is only capable of identifying subgroups within data, and lacks the ability to infer the shared beliefs within these groups.

Despite the fact that a clear understanding of public attitudes towards AI is critical for its successful integration and acceptance in society, little research has been conducted to investigate the consensus beliefs within society towards responsible AI. To fill this gap, the current study expands on the CCT and applies it to the public's perception of responsible AI. The study utilizes a nationally representative U.S. sample and surveys them about up-to-date AI applications, aiming to provide valuable insights to developers and policymakers.

\section{Cultural Consensus Theory}
The underlying principle of consensus-based models is that, for many tasks, the group's central tendency often yields a precise outcome. The collective response can serve as a substitute for the actual answer when assessing individual group members—those whose judgments are closer to the group's central tendency (across multiple questions) are presumed to possess greater knowledge. Consequently, consensus-based models can be employed to estimate an individual's level of knowledgeability when there is no available ground truth. Cultural Consensus Theory (CCT), developed by Romney et al. \cite{romney1986culture}, is a theory and method that outlines the conditions under which agreement among individuals can indicate knowledge or accuracy. Many folk epistemological systems rely on the relationship between agreement and truth, such as the court jury system and the scientific practice of taking multiple measurements and averaging them. The idea of collective wisdom is also prevalent in influential works like ``The Wisdom of Crowds'' \cite{surowiecki2005wisdom} and in common phrases like ``50,000,000 Elvis fans can't be wrong.'' However, it is evident that agreement does not always equate to accuracy. The theory of consensus analysis identifies the specific circumstances in which agreement can indeed imply knowledge.

Researchers employing CCT strive to measure the consensus from respondents' individual responses, where the researchers do not know the consensus ahead of time, nor do they know which respondents have more or less knowledge. Indeed, the task of CCT is akin to simultaneously determining an answer key for a test given to test-takers and scoring the ability of those test-takers with respect to the answer key \cite{batchelder1988test, oravecz2015hierarchical}. Additionally, as with related Item Response Theory (IRT) methods, CCT measures the difficulty of the questions. In CCT, these tasks are all accomplished with cognitive-based response process models, with consensus answers and cognitive characteristics of the respondents estimated endogenously. Thus, CCT is useful to researchers in situations where (a) the consensus knowledge of the participants is unknown to the researcher ahead of time, (b) the researcher has access to a limited number of participants who may or may not have had equal access to this shared cultural knowledge, (c) the researcher can construct a relevant questionnaire but does not know which questions are more or less difficult, and (d) the researcher does not know much about the characteristics of the participants respondents \cite{romney1986culture}.

In the current study, we use a Continuous Response Model (\textsc{crm}), developed by Anders et al. \cite{anders2014cultural} that allows for multiple consensus truths, which serves as the basis for our extension to the model.

\subsection{Infinite Cultural Consensus Theory}

CCT has been used to infer the cultural beliefs that influence social practices and assess the extent to which individuals comprehend and manifest these beliefs \cite{romney1986culture}. In instances of multiple cultures, CCT applies the analysis of eigenvalues derived from the cross-participant correlation matrix to determine the number of cultures present. However, this approach encounters two main challenges: it presumes a finite-dimensional representation that accurately characterizes the features of observed data and assumes minimal missing values in the observed data.

In response to these issues, we introduce a Bayesian non-parametric model in which cultures are derived from a Dirichlet Process (DP) through a stick-breaking construction method. This method offers a fully probabilistic approach to inferring the existing number of cultures that contribute to individuals' judgment of AI. In this non-parametric approach, the objective is to learn from data without imposing any rigid assumptions about the class of distributions that may describe the data. The reasoning behind this method is the acknowledgement that the generative process for data is unlikely to conform to any finite-dimensional distribution family, thus avoiding potential incorrect assumptions.

The DP provides solutions to Bayesian non-parametric inferences by utilizing a prior over discrete distributions. This technique is extensively employed in Bayesian nonparametric models as an infinite discrete prior for mixture models \cite{griffiths2003hierarchical}, and has been applied to psychometric models as a prior over probability \cite{duncan2008nonparametric}.

\section{Methods}
\subsection{Data collection}

The data used in this research was gathered from a survey conducted by Morning Consult \cite{moriningconsult} between September 8 and September 10, 2021. The sample consisted of 2105 adults, and the data collection was carried out online. Morning Consult uses a methodology that involves collecting data from large groups of respondents and then weighting that data to ensure it is representative of the studied population in terms of demographic features, such as age, gender, race, and education. During data collection, Morning Consult compares the demographic makeup of their sample to that of the studied population. For instance, if the U.S. population is 51\% female, disregarding other demographic features, the collected data should reflect a similar portion of female respondents. This allows them to gather data that is both diverse and representative, enhancing the validity and generalizability of the findings. 

\subsection{Questionnaires}

We formulated the survey questions to capture the major facets of responsible AI. \textbf{Q1} explores public concerns about negative AI outcomes. \textbf{Q2} examines public optimism regarding AI's benefits. \textbf{Q3} evaluates public views on the ethics of AI applications. \textbf{Q4} assesses perceptions of AI governance and oversight. \textbf{Q5} probes trust in AI's regulatory bodies. \textbf{Q6} measures public perception of AI's societal impact. These six main questions, each with a number of specific focus points, culminating in a total of 62 questions. We provide the main questions along with the list of specific focuses below. The response options used an ordinal scale, ranging from 1 to 4, where 1 represented the most negative response, and 4 represented the most positive response. The specific wording for the responses varied according to the question. We rescaled the ratings to fall within the interval (0, 1) because CCT links the random response variables in this range to the real line using the logit transformation.

\textbf{Q1.} How concerning, if at all, are each of the following potential negative consequences of greater AI adoption in everyday life? \emph{Less personal interaction, reduced employment opportunities, racial/ethnic bias in AI, gender bias in AI, overreliance on AI, reduced creativity, lack of empathy in AI, reduced autonomy in human decision making, reduced human connectedness, AI gaining consciousness, AI becoming uncontrollable, lack of transparency in AI decision-making, irresponsible use of AI among countries or businesses, lack of understanding about how AI works, loss of personal privacy, increased spread of misinformation, increased political polarization, increased global conflicts.}

\textbf{Q2.} How likely or unlikely is it that each of the following will be potential positive outcomes of greater AI adoption in everyday life? \emph{Increased economic prosperity, increased innovation, more consistent decision making, smarter technology, improved technology efficiency, reduced human error, more widespread use of technology, reduced risk to humans in
dangerous jobs, 24/7 availability, ability to handle repetitive tasks, increased job creation in the
technology field, better working conditions}

\textbf{Q3.} How responsible or irresponsible are each of the following current or proposed uses of
artificial intelligence? \emph{Self-driving cars, facial recognition, engaging in military conflict, deepfakes, service industry job automation, lip reading and speech recognition, manufacturing job automation, search engine and personal shopping optimization, healthcare automation, personal assistants, directing air traffic, managing personal finances, delivering packages, answering 911 calls, testing self-driving car technologies on public roads.}

\textbf{Q4.} In your opinion, how well regulated is artificial intelligence at the moment?

\textbf{Q5.} Do you trust or distrust each of the following entities to develop and oversee the standards that should be used to regulate the use of artificial intelligence in public life, either through legislation governing its use or a moral and ethical set of professional standards? \emph{Federal government, state governments, local governments, international bodies, companies that develop AI, companies that use AI, colleges and universities, professional societies, Standards organizations, foundations and non-profits.}

\textbf{Q6.} How much of a role, if at all, has AI played in each of the following in the U.S.? \emph{Loss of trust in elections, threats to democracy, distrust in vaccines, loss of trust in institutions, political polarization, spread of misinformation}

\section{Results}
\subsection{Consensus beliefs regarding responsible AI}
Our  iCCT model revealed that a representative sample of the U.S. population includes four consensus beliefs concerning various aspects of responsible AI. These findings effectively address our first research question, illustrating that the U.S. population does not possess uniform beliefs about responsible AI (Fig. 1 and Fig. 2). The diverse perspectives emphasize the complexity and nuanced understanding of artificial intelligence within the U.S. populace. Furthermore, it indicates that our society's discourse around responsible AI is multi-faceted and influenced by a variety of factors such as individual differences and cultural constructs.

While the number of cultures indicates the number of consensus in the data, it does not inform us about the uniformity of the cultural assignment distribution. An entropy-based metric addresses this problem by estimating the uncertainty of the cultural allocation of an unknown randomly chosen data point given a particular distribution of culture assignments. The smaller value of entropy indicates that there are a few large clusters, and the larger values of entropy are associated with more evenly distributed cultures. In the analysis, the final cultural assignment is determined by the modal membership across the final 100 posterior samples. Out of 2105 respondents, 1217 respondents were clustered into culture 1, 504 were clustered into culture 2, 218 were clustered into culture 3, and 166  were clustered into culture 4. This can be attributed to the presence of one large cluster (cluster 1, holding approximately 55.3\% of the data points), while the remaining clusters are smaller in comparison. The cluster entropy is not exceedingly low due to a moderate level of uniformity observed among the three smaller clusters.

\renewcommand{\thefigure}{1}
\begin{figure}[h]
\centering
\includegraphics[width=5.5cm, height=5.5cm]{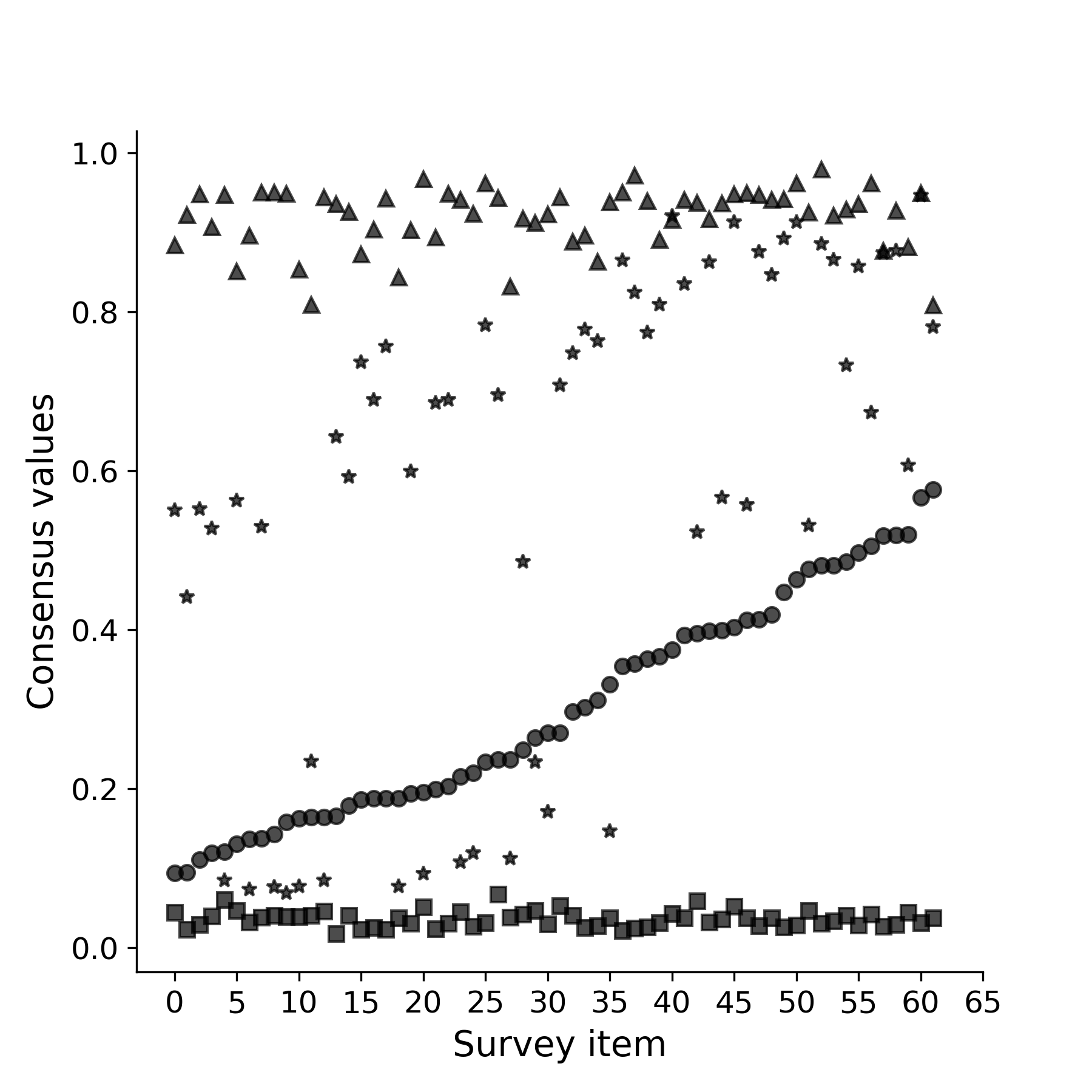}
\caption{Cultural cluster's posterior mean results of consensus values from the responsible AI survey items. Circles (culture 1), stars (culture 2), squares (culture 3), and triangles (culture 4) respectively represent the 'truths' of the four identified cultures.}
\end{figure}

\renewcommand{\thefigure}{2}
\begin{figure*}[h]
\centering
\includegraphics[width=0.55\textwidth]{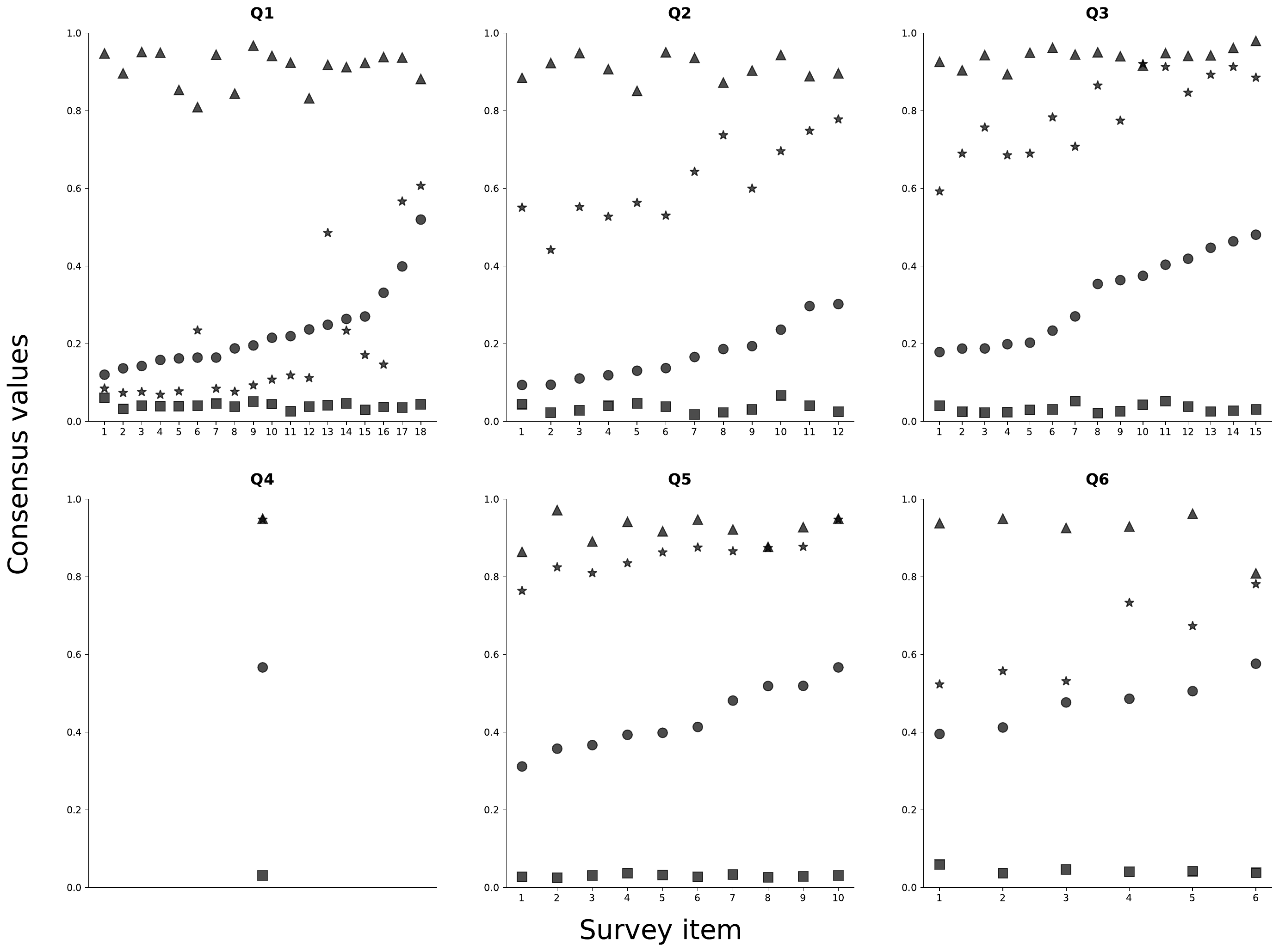}
\caption{Cultural clusters' posterior mean results for each main question. Circles, stars, squares, and triangles respectively represent the 'truths' of the four identified cultures.}
\end{figure*}

Fig. 1 presents the posterior mean inverse logit item consensuses for each of the four identified cultures. It appears that respondents assigned to the 'triangle' culture consistently exhibit more positive sentiments towards responsible AI compared to those from other cultures. Conversely, those assigned to the 'square' culture demonstrate consistently more negative attitudes towards responsible AI compared to counterparts from the other cultures.

Fig. 2 illustrates the posterior mean inverse logit item consensuses of each main question for each of the four cultures. Generally speaking, the circle, star, and square cultures exhibit substantial concern regarding potential negative consequences of greater AI adoption. Although the star culture shows concern about the negative consequences of AI, they believe AI is well regulated and used responsibly, and they indicate a high level of trust in entities that are developing and overseeing the standards that should be used to regulate the use of AI in public life. 

While our findings confirm that the U.S. population holds diverse consensus beliefs about responsible AI, it is important to note that different cultures exhibit similar consensus beliefs for the question \textbf{(Q1)} addressing the \emph{'lack of transparency in AI decision-making'}. Excluding culture 4 (represented by a triangle in Fig. 1), three of the cultures expressed a strong concern over this lack of transparency in AI decision-making processes. This commonality suggests a shared perception across these cultures about how AI decisions are made and the potential impacts of these decisions remaining opaque. As such, it is clear that the issue of transparency is a significant facet of the broader conversation about responsible AI, deserving of close attention in both technological development and policy-making discussions.

\subsection{Controversial and challenging topics in responsible AI: Inter and intra-cultural consensus beliefs}

Compared to previous methods, such as segmentation analysis and topic modeling, iCCT is able to identify subjects related to responsible AI where cultural disagreements are most prominent. This is accomplished by focusing on items exhibiting the highest variance among consensus values. In our analysis, we found the consensus beliefs for \textbf{Q3}, which focuses on ``\emph{manufacturing job automation}'', exhibited the highest variation among the cultures, with respective consensus values of .18, .75, .02, and .94. Automation of manufacturing jobs through the use of artificial intelligence is indeed a controversial topic (Wang, 2019). While the prospect of automation in the manufacturing sector poses a threat to job security for a significant number of workers worldwide, some individuals may perceive it as a means to increase efficiency and productivity in manufacturing.

\begin{flushleft}
\begin{table}
\caption{Consensus values for each culture across the most controversial aspects of main questions, along with their mean consensus values}
\centering
\resizebox{0.475\textwidth}{!}{%
\begin{tblr}{
  colspec={ccccc|}, 
  cell{1}{2} = {c=5}{c}, 
  cell{2}{2} = {c},
  cell{2}{3} = {c},
  cell{2}{4} = {c},
  cell{2}{5} = {c},
  cell{2}{6} = {c}, 
  row{1} = {font=\bfseries}, 
}
                                                    & \textbf{Cultures} &            &            &            &  \\
                                                    & \textbf{1}        & \textbf{2} & \textbf{3} & \textbf{4} & \textbf{Mean}              \\
\textbf{Q1 -}~\emph{Less personal interaction}    & .30               & .77        & .01        & .89        & .34 \\
\textbf{Q2 -}~\emph{Smarter technology}           & .11               & .55        & .03        & .93        & .43 \\
\textbf{Q3 -}~\emph{Manufacturing job automation} & .18               & .75        & .02        & .94        & .52 \\
\textbf{Q5 -}~\emph{Colleges and universities}    & .35               & .82        & .02        & .97        & .56 \\
\textbf{Q6 -}~\emph{Threats to democracy}         & .50               & .67        & .04        & .96        & .52        
\end{tblr}
}
\end{table}
\end{flushleft}

Table 1 presents the most controversial aspects of each question across the cultures, revealing that there are subgroups within the population holding different perspectives on 'smarter technology' as a potentially detrimental outcome of AI adoption in everyday life. This disparity could account for the nuanced viewpoints regarding the societal implications of AI technologies. In this context, 'smarter technology' not only refers to improvements in efficiency and productivity, but also to a broader transformation of societal norms and structures. For instance, these differing attitudes might reflect underlying concerns or optimism about AI's role in reducing personal interactions, reshaping manufacturing jobs through automation, and posing potential threats to democracy. Some subgroups may perceive these changes as an advancement, while others might view them as a loss or a risk. Understanding these multifaceted views is key to fostering a balanced and inclusive dialogue on the future of AI. These results adequately address the second research question.

Views on trusting 'higher education institutions' to regulate the use of artificial intelligence greatly vary among different subgroups within the U.S. population. This divergence may arise from the fact that a large portion of AI developments are led by major tech companies rather than academic institutions, and these companies often maintain a robust media presence, shaping public opinion and perceptions. Furthermore, the level of trust individuals and groups place in colleges and universities to regulate AI can vary significantly, often influenced by the perceived competency of these institutions in the field of AI. This perception contributes to the divergent views regarding their role in AI governance.

In order to answer our third research question, identifying the controversial issues and aspects related to responsible AI can assist developers and policymakers in formulating interventions to mitigate any negative impacts on public attitudes towards responsible AI. Moreover, pinpointing the challenging issues and facets of responsible AI within specific subgroups can offer valuable insights. By focusing on these particular topics, developers and policymakers can potentially prevent further divergence in viewpoints and concentrate on addressing specific concerns. In iCCT, the item difficulty parameter is conceptualized as a function of the intrinsic characteristics of a question. It can be determined by factors such as the complexity of the topic, the level of knowledge needed to answer it cultural correct, or the degree of ambiguity or contention associated with the issue within the cultural context.

\renewcommand{\thefigure}{3}
\begin{figure}[h]
\centering
\includegraphics[width=5cm, height=5cm]{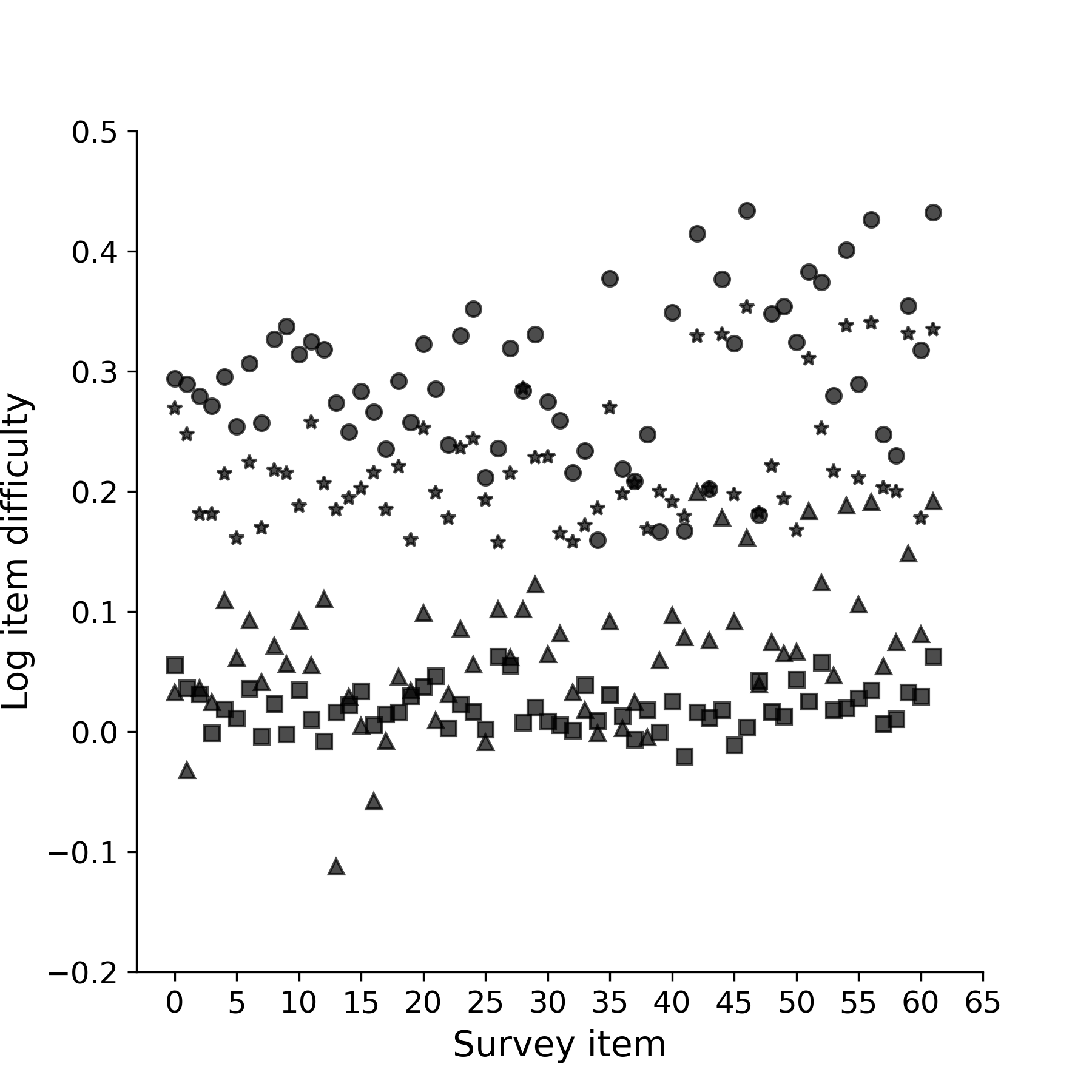}
\caption{Posterior mean results of the item difficulties for each culture.}
\end{figure}

The item difficulty parameter is valuable for identifying questions that may necessitate more focused communication efforts, particularly when the objective is fostering a shared understanding within a group. Figure 3 demonstrates which items each culture found the most challenging to assess accurately. Cultures 1 and 2 identified the question about the role of AI in the 'loss of trust in elections' in the United States as the most challenging to answer. Culture 3 found the question about the influence of AI on the 'spread of misinformation' to be the most difficult. Culture 4 found it most challenging to assess the likelihood of 'more consistent decision making' as a potential positive outcome of increased AI adoption in everyday life.

\renewcommand{\thefigure}{4}
\begin{figure}[h]
\centering
\includegraphics[width=5cm, height=9cm]{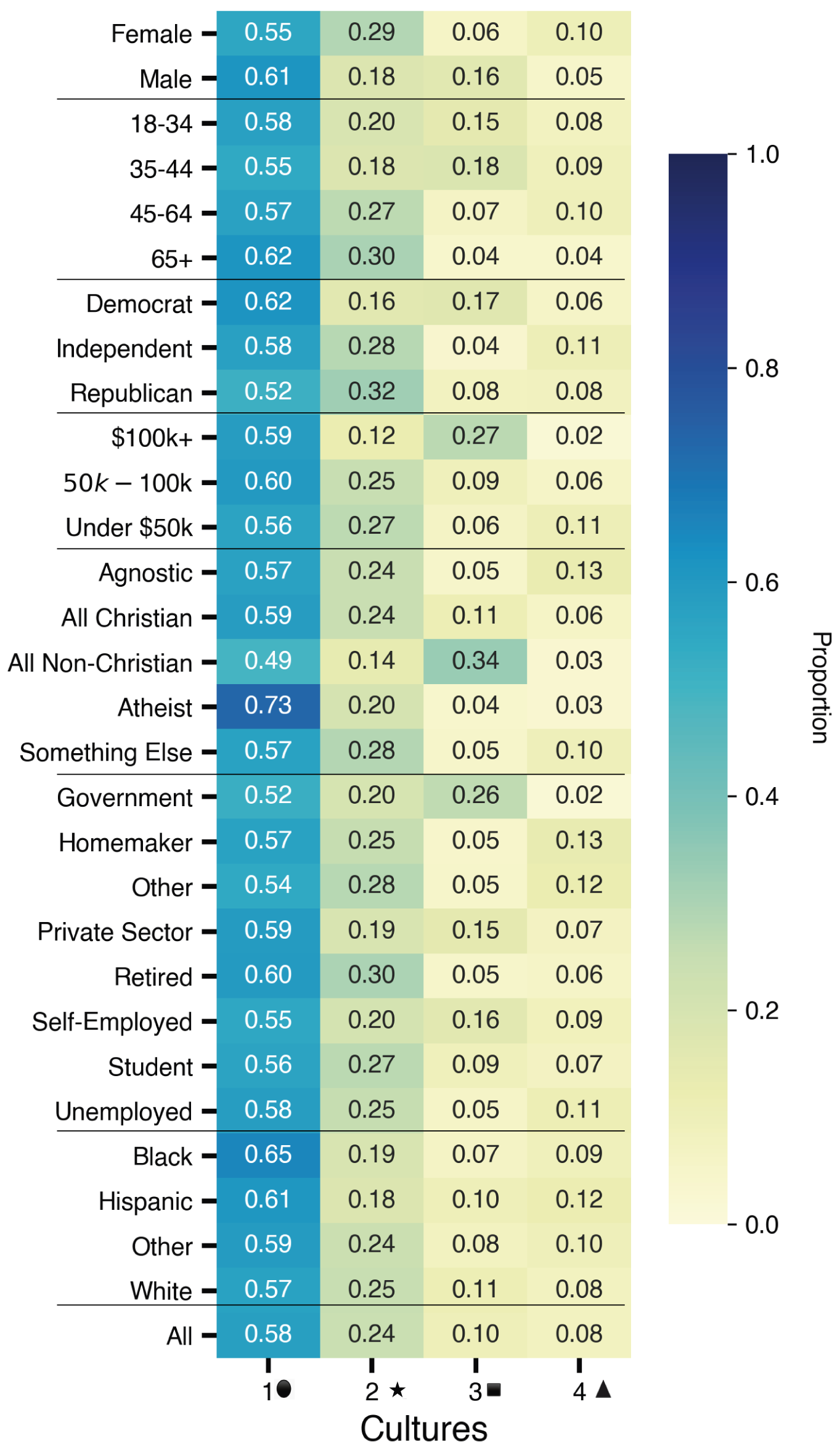}
\caption{Demographic Distribution of Cultural Assignments}
\end{figure}

As depicted in Fig. 1 and Fig. 2, people's perceptions of responsible AI include a wide spectrum of attitudes and beliefs, rather than presenting a uniform perspective. Interestingly, these variations do not appear to be directly correlated with geographic and demographic features like age, gender, or location (Fig. 4 and 5). In the interconnected society we live in, people's perceptions are influenced by a complex network of factors that extend beyond basic geographic and demographic attributes. A significant part of this complexity is contributed by the easily accessible and diverse array of viewpoints on AI found on social media, online platforms, and professional and social networks. These platforms not only allow individuals to form unique understandings but also expose them to a variety of perspectives. Fig. 4 adds a layer of nuance by indicating a slight trend of negative perceptions of responsible AI among individuals working in government, earning over \$100k, and identifying as non-Christian. However, Fig. 4 also emphasizes that demographic features alone do not provide a clear, predictable insight into people's perceptions of responsible AI. This insight illuminates the multifaceted and nuanced nature of responsible AI perceptions in our interconnected society, highlighting that these perceptions are not just products of individual demographic attributes, but also of the shared knowledge and perspectives within our interconnected social fabric.

To determine whether the identified cultural clusters correlate with the geographic location of participants, we used the participants' ZIP codes to map their locations. As depicted in Fig. 5, people's perception of responsible AI cannot be explained by their geographical location.This conclusion aligns with the insights derived from demographic features. That is, perceptions do not merely originate from individual geographic characteristics, but also embody the collective knowledge and viewpoints prevalent in our intricately interconnected societal structure.

\renewcommand{\thefigure}{5}
\begin{figure}[h]
\caption{Geographic Distribution of Cultural Assignments}
\centering
\includegraphics[width=7cm, height=5.5cm]{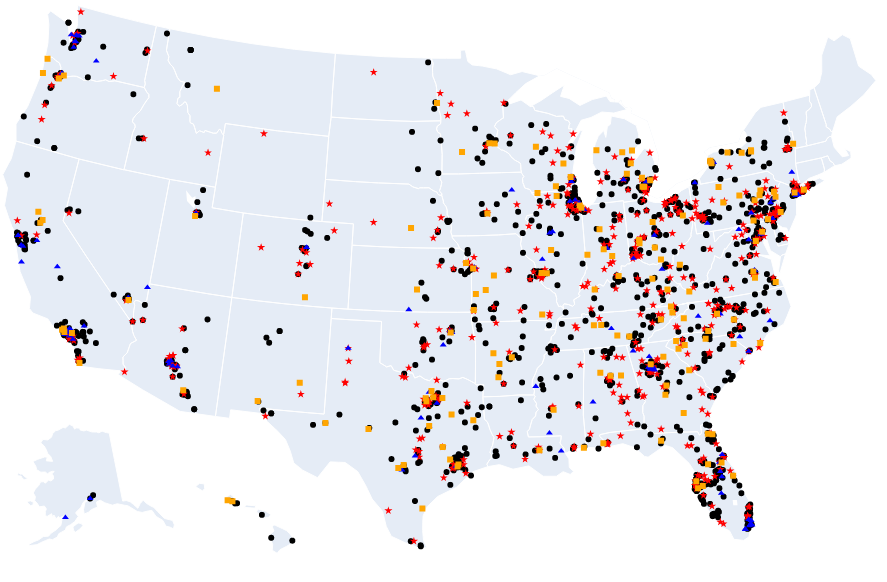}
\end{figure}

\section{Discussion}

This study investigated consensus beliefs regarding responsible AI using nationally representative survey data, yielding valuable insights for developers and policymakers. By extending the Cultural Consensus Theory (CCT) through a Bayesian non-parametric process, we demonstrated that the U.S. population holds various attitudes towards responsible AI. Our findings revealed four distinct consensus beliefs regarding various aspects of responsible AI within the population, with one large cluster and three smaller ones. The cluster entropy validates a moderate level of uniformity among the three smaller clusters. These insights, thus, emphasize the importance of tailored communication strategies that consider the distinctive beliefs within each cluster to encourage more informed discourse and decision-making regarding responsible AI.

While recent research has begun to examine the public's attitudes towards various aspects of AI \cite{bao2022whose, jakesch2022different}, these studies have not been able to extract the consensus beliefs of each subgroup and have overlooked individual and group-level differences. By identifying different consensus beliefs among population subgroups, we bring attention to the most controversial topics and aspects of various AI applications, and their impact on public perception. The most controversial issue among the subgroups was the responsible use of AI in automating manufacturing jobs. This discovery highlights the need for policymakers and developers to invest more time in mitigating these divergent concerns to enhance the public's perception of responsible AI, which is crucial for the successful adoption of such technologies. Therefore, this research reemphasizes the importance of a nuanced, stakeholder-oriented approach in the development and regulation of AI, ensuring that the technology progresses in a manner that respects societal norms and expectations.

While demog raphic and geographic attributes of participants have provided some degree of explanation for the observed variations in perceptions regarding responsible AI, they do not fully and robustly account for the complexity of these perceptions. Our findings indicate that a more nuanced approach may be necessary to better understand this complex phenomenon. Researchers might consider employing more sophisticated methodologies, such as cognitive response analyses or psychometric models. These methodologies can offer deeper insights by dissecting the survey data at a more granular level, enabling the examination of the subtle cognitive processes and psychological metrics that might underlie and contribute to people's attitudes towards responsible AI. By embracing such intricate approaches, we can gain a more comprehensive understanding of how various factors interplay to shape individual and collective perceptions of responsible AI.

Despite contributing valuable insights into the public's view of responsible AI, our study was inherently limited due to the constrained scope of six main questions. Future research could provide a more holistic understanding by implementing a broader and more detailed questionnaire, exploring the finer aspects of AI responsibility. By incorporating this expanded range and considering emerging trends in AI, we can create more refined strategies for the responsible development and implementation of AI. Moreover, future studies can incorporate qualitative (e.g., interviews) with quantitative methods to deepen our understanding of public's opinion of responsible AI.


\bibliographystyle{unsrt}  
\bibliography{references}  


\end{document}